\documentclass[11pt, a4paper]{article}
\usepackage[margin=2.5cm]{geometry}
\usepackage[utf8]{inputenc}
\usepackage{authblk}
\usepackage{graphicx}
\usepackage{subcaption}
\usepackage{amsmath}
\DeclareMathOperator*{\argmax}{arg\,max} 

\title{Locating the source of interacting signal in complex networks}
\author[1]{R. Paluch}
\author[1]{K. Suchecki}
\author[1,2]{J. A. Hołyst}

\affil[1]{Faculty of Physics, Warsaw University of Technology, Koszykowa 75, 00-662 Warszawa, Poland}
\affil[2]{ITMO University, Kronverkskiy Prospekt 49, St Petersburg, Russia 197101}

\begin{document}

\maketitle

\begin{abstract}
We investigate the problem of locating the source of a self-interacting signal spreading in a complex networks.
We use a well-known rumour model as an example of the process with self-interaction.
According to this model based on the SIR epidemic dynamics, the infected nodes may interact and discourage each other from gossiping with probability $\alpha$.
We compare three algorithms of source localization: Limited Pinto-Thiran-Vettarli (LPTV), Gradient Maximum Likelihood (GMLA) and one based on Pearson correlation between time and distance.
The results of numerical simulations show that additional interactions between infected nodes decrease the quality of LPTV and Pearson.
GMLA is the most resistant to harmful effects of the self-interactions, which is especially visible for medium and high level of stochasticity of the process, when spreading rate is below 0.5.
The reason for this may be the fact that GMLA uses only the nearest observers, which are much less likely affected by the interactions between infected nodes, because these contacts become important as the epidemics develops and the number of infected agents increases.
\end{abstract}

\section{Introduction}
\label{section-introduction}

Studies of information spreading are an active and important research field and it comes as no surprise, given the fundamental role of information in the present society.
In fact there are many attempts to measure and model  \cite{LibenNowell2008,Moreno2004,Nekovee2007,Miritello2011,Pei2013} these phenomena.
Often it is  evident what the source of information is, e.g. for  messages carrying official content  and consistently  citing information source.
A specific issue of locating an unknown information source received significant attention and several methods have been developed, from based on single-instance snapshot \cite{Shah2011, Prakash2012, Lokhov2014, Zhu2016b} to observer-based approaches where specific nodes are observed through the process \cite{Pinto2012, Karamchandani2013, Luo2014, Paluch2018, Xu2019}.
In most studies attempting to locate the source, simple spreading models -- SI or SIR have been applied.
While using these  universal models allows to apply conclusions not only for  knowledge dissemination but also for infectious diseases spread, the issue of attempting to locate the source of information spreading according to a more complex  dynamics  has  not been investigated.
It is a challenge to   develop methods being able to find an unknown source  in the case when  information does not spread openly, for example in the form  of fake news, conspiracy theories or rumours.
Developing such methods is of primary importance  for society because these types of messages  are often detrimental and may even pose danger, such as false anti-vaccine rumours or conspiracy theories, and locating a primary source may help in preventing their spread in the future.

In this work we have investigated a model specifically developed to represent rumour spreading behavior \cite{Nekovee2007}, where  spreaders may influence each other.
Such self-interacting signals may exhibit properties and behaviors that regular non-interacting signals do not possess, and which interfere with attempts to locate the original spread sources.
We have investigated how methods developed to locate source of spreading information can cope with the challenge of more specific, self-interacting rumours.

We introduce our rumour model in section \ref{section-model}, in section \ref{section-methods} we present  methods used to locate a rumour source, in section \ref{section-results} we show results of our methods  and in section \ref{section-discussion} we discuss our observations.

\section{Rumour model}
\label{section-model}
In this work we use the agent-based version of popular rumour model proposed by Moreno and Nekovee \cite{Moreno2004, Nekovee2007}.
This model in turn can be understood as an extension of the classical SIR model \cite{Kermack1927}, therefore the names of agents' states in our paper refer to the SIR model.
An agent may be in one of three states during the simulation:
\begin{itemize}
    \item Susceptible (S) -- an agent did not hear a rumour or is not interested in gossiping. This state is also called \textit{ignorant} in other publications.
    \item Infected (I) -- an agent spreads the rumour, which means that it can affect the state of his susceptible or infected neighbours. This state is also called \textit{spreader}.
    \item Recovered (R) -- an agent stops gossiping and starts persuading their infected neighbors to do the same. Another name for this state is \textit{stifler}.
\end{itemize}

The model allows for the following interactions:
\begin{itemize}
    \item $I + S \overset{\beta}{\to} I + I$ -- an infected agent has a probability $\beta$ per time step to infect its susceptible neighbour.
    \item $I + I \overset{\alpha}{\to} I + R$ -- when an infected agent tells the rumour to the infected neighbour, it can unconsciously discourage the neighbour for further gossiping with a probability $\alpha$.
    \item $R + I \overset{\alpha}{\to} R + R$ -- a recovered agent has a probability $\alpha$ per time step to persuade its neighbour to stop gossiping.
\end{itemize}

\begin{figure}[!hbt]
    \centering
    \includegraphics[width=\textwidth]{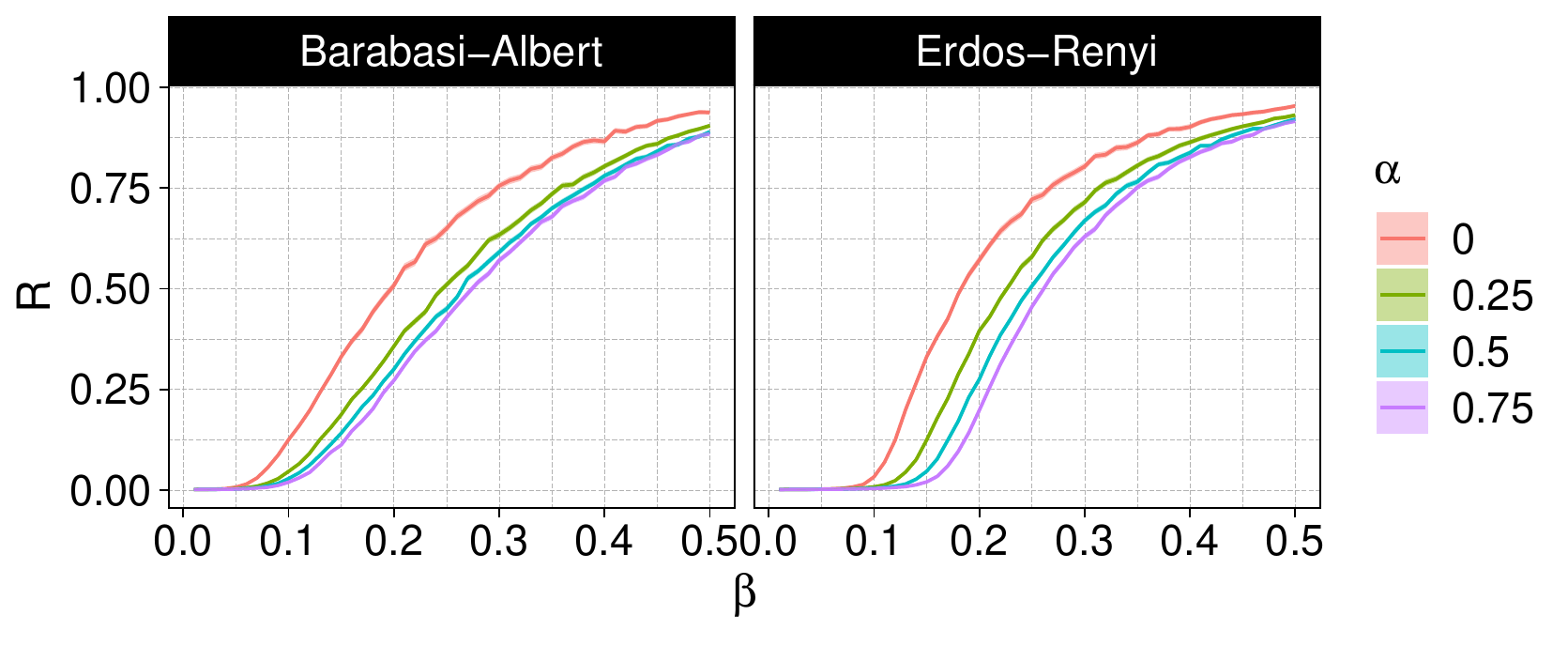}
    \caption{Final rumour size in Barab\'{a}si-Albert and Erd\H{o}s–R\'{e}nyi graphs with $n=1000$ and $\langle k \rangle = 6$. 
    Forgetting rate $\gamma = 0.5$.
    The results are averaged over $5\cdot10^3$ runs.}
    \label{fig:R_beta}
\end{figure}

The dynamics of our model is synchronous, which means that at every time step, all infected nodes pass the rumour simultaneously to their all neighbours (which may change susceptible neighbours into infected and infected into recovered with corresponding probabilities $\beta$ and $\alpha$).
Similarly, in the same moment each recovered node tries to convince its infected neighbour to stop gossiping with a probability $\alpha$.
Regardless of the interactions performed, at the end of a time step each infected agent may become recovered with a probability $\gamma$.
This mechanism corresponds to losing the interest for the rumour or simply forgetting it. 
The main difference between SIR and rumour model is the self-interaction of the signal resulting from the fact that an infected agent may stop gossiping after hearing the rumour again from the neighbour.
This mechanism reflects the psychological effect that people are more likely to gossip about hot new topics.
As shown in Fig. \ref{fig:R_beta}, the internal interaction of signal, tuned by the parameter $\alpha$, decreases the final rumour size in the system, which is the fraction of recovered nodes at the end of epidemics (when there is no infected agents left).

\section{Localization algorithms}
\label{section-methods}
To investigate the impact of the self-interaction of signal on the quality of the source localization we use three detectors-based inference algorithms: Pinto-Thiran-Vetterli \cite{Pinto2012}, Gradient Maximum Likelihood Algorithm \cite{Paluch2018} and Pearson correlation \cite{Xu2019}.
The common feature of these methods is that they use detectors (also called observers) placed in the nodes of network to measure the times when the signal arrived to this nodes.
Then the information from the observers is used for computation a \textit{score} for each node $s$, denoted as $\mathcal{F}(s)$.
The estimated source $\hat{s}$ of the rumour is given by:
\begin{equation}
    \hat{s} = \argmax \mathcal{F}(s)
    \label{eq:mle}
\end{equation}

Moreover, all methods mentioned above require the full knowledge about the topology of graph, since they need to find the shortest paths between node $s$ and all observers.
The next paragraphs briefly discuss the algorithms.

\subsection{Pinto-Thiran-Vetterli}
The central part of algorithm proposed by Pinto et al. \cite{Pinto2012} is the maximum likelihood estimator derived for trees (acyclic undirected graphs) with the assumption about gaussian distribution of time delays between nodes.
To apply the method for generic graphs, one should leave only the shortest paths between node $s$ and all observers.
Let $|\mathcal{P}(u,v)|$ denote the length of the path $\mathcal{P}(u,v)$ connecting nodes $u$ and $v$, $K$ is the number of observers, ${d}$ the vector of observed delays, ${\mu}_s$ the vector of deterministic delays and ${\Lambda}_s$ the delay covariance matrix, then the score is:

\begin{equation}
\mathcal{F}(s) = \mu_{s}^{T} \Lambda^{-1} (d - \frac{1}{2}\mu_s),
\label{eq:ptv_score}
\end{equation}
where:
\begin{equation}
\begin{aligned}
\lbrack{d}\rbrack_k &= t_k - t_0, \\
\lbrack{\mu}_s\rbrack_k &= \mu (|\mathcal{P}(s,o_k)|-|\mathcal{P}(s,o_0)|),\\
\lbrack{\Lambda}_s\rbrack_{k,i} &= \sigma^2 |\mathcal{P}(o_0,o_k)\cap\mathcal{P}(o_0,o_i)|, \;\; for \;k,i=1,\dots,K-1.
\end{aligned}
\end{equation}

In the original work of Pinto et al., the observers also know from who they got infection and use this knowledge for the source locating.
In this work we do not use this information and therefore we call this method Limited Pinto-Thiran-Vetterli (LPTV).

\subsection{Gradient Maximum Likelihood Algorithm}
This method is built on the same maximum likelihood estimator as LPTV, but it underlines the meaning of the observers which are the closest to the source.
GMLA has an additional parameter $K_0$, which is the number of the nearest observers (with the lowest delays) used by the algorithm.
Typically $K_0 \ll K$, which reduces significantly the time of the score computation.
Moreover, GMLA do not compute the score for each node in the graph but it performs gradient-based selection of suspected nodes.
These to modification accelerate greatly the process of the source localization without a notable reduction in precision.

\begin{figure}[!hbt]
    \centering
    \includegraphics[width=\textwidth]{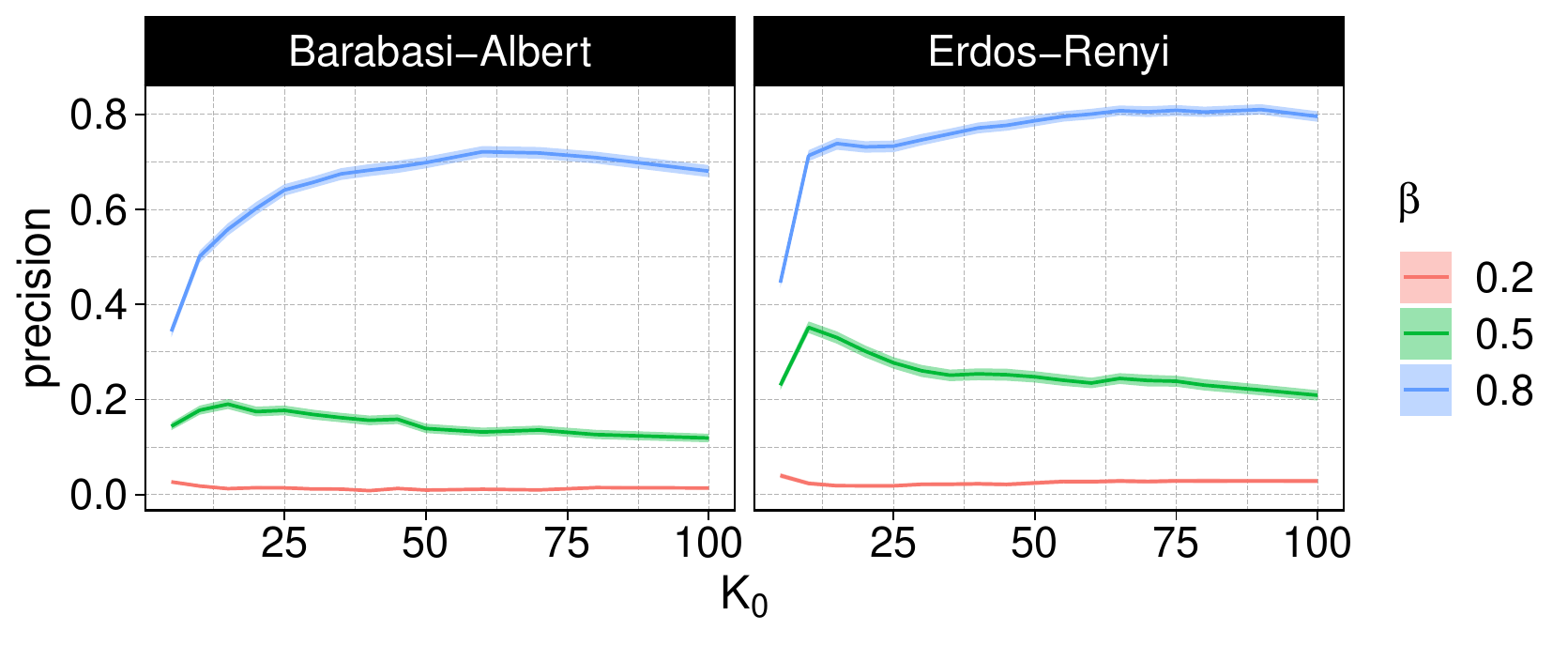}
    \caption{Average precision of GMLA in Barab\'{a}si-Albert and Erd\H{o}s–R\'{e}nyi graphs with $n=1000$ and $\langle k \rangle = 8$ versus the number of nearest observers $K_0$.
    Density of observers $\rho = 0.1$.
    Forgetting rate $\gamma = 0.0$.
    The results are averaged over $5\cdot10^3$ runs.}
    \label{fig:precision_vs_no}
\end{figure}

In order to tune the appropriate value of $K_0$ we conduct the experiment for different values of the spreading rate $\beta$.
Figure \ref{fig:precision_vs_no} shows that the optimal number of the nearest observers is similar for Barab\'{a}si-Albert and Erd\H{o}s–R\'{e}nyi and increases with $\beta$.
This means that for more deterministic signal (higher $\beta$) even observers which are far from the source may be useful in the localization process.
For $\beta = \{0.2, 0.5, 0.8\}$ we select $K_0^{BA} = \{5, 15, 60\}$ and $K_0^{ER} = \{5, 10, 60\}$ which we use in the studies presented in section \ref{section-results}.

\subsection{Pearson correlation}
The algorithm of Xu et al. \cite{Xu2019} is based on a simple remark that the order in which the observers are infected should be consistent with the order of the distances between the observers and the source.
The authors measure this dependency using Pearson correlation:
\begin{equation}
    \mathcal{F}(s) = \frac{\sum_{i=0}^{K-1}([d_s]_i - \overline{d_s})(t_i - \overline{t}) }
    { \sqrt{ \sum_{i=0}^{K-1} ([d_s]_i - \overline{d_s})^2  \sum_{i=0}^{K-1}(t_i - \overline{t})^2 } }
\end{equation}
where $[d_s]_i$ is the shortest path length between the node $s$ and the observer $i$, $K$ is the number of observers and $t_i$ is the time of infection of observer $i$.

\section{Results}
\label{section-results}
To study how the self-interaction of the spreading process affects the quality of source detection we conduct series of numerical experiments on two types of network, Erd\H{o}s–R\'{e}nyi graph and Barab\'{a}si-Albert model.

We use two efficiency measures for evaluating the quality of source detection: the average precision and the Credible Set Size at $0.95$ confidence level.
The precision for a single test is defined as the ratio between true positives and the sum of true positives and false positives sources.
The tests are repeated multiple times (typically $10^3$) and then the average value of precision is computed.
The Credible Set Size at the confidence level of $\alpha$ ($\alpha\text{-CSS}$) \cite{Paluch2020} is the size of the smallest set of nodes containing the true source with probability $\alpha$.
In order to estimate $\alpha\text{-CSS}$ the following procedure is performed.
First, the multiple tests of source detection are conducted and for each test the rank of the true source is computed.
The rank is the position of the node on a list in descending order of the nodes' scores.
Next, $\alpha\text{-CSS}$ is computed as a $\alpha\text{-quantile}$ of the source rank. 

\begin{figure}[!hbt]
    \centering
    \includegraphics[width=\textwidth]{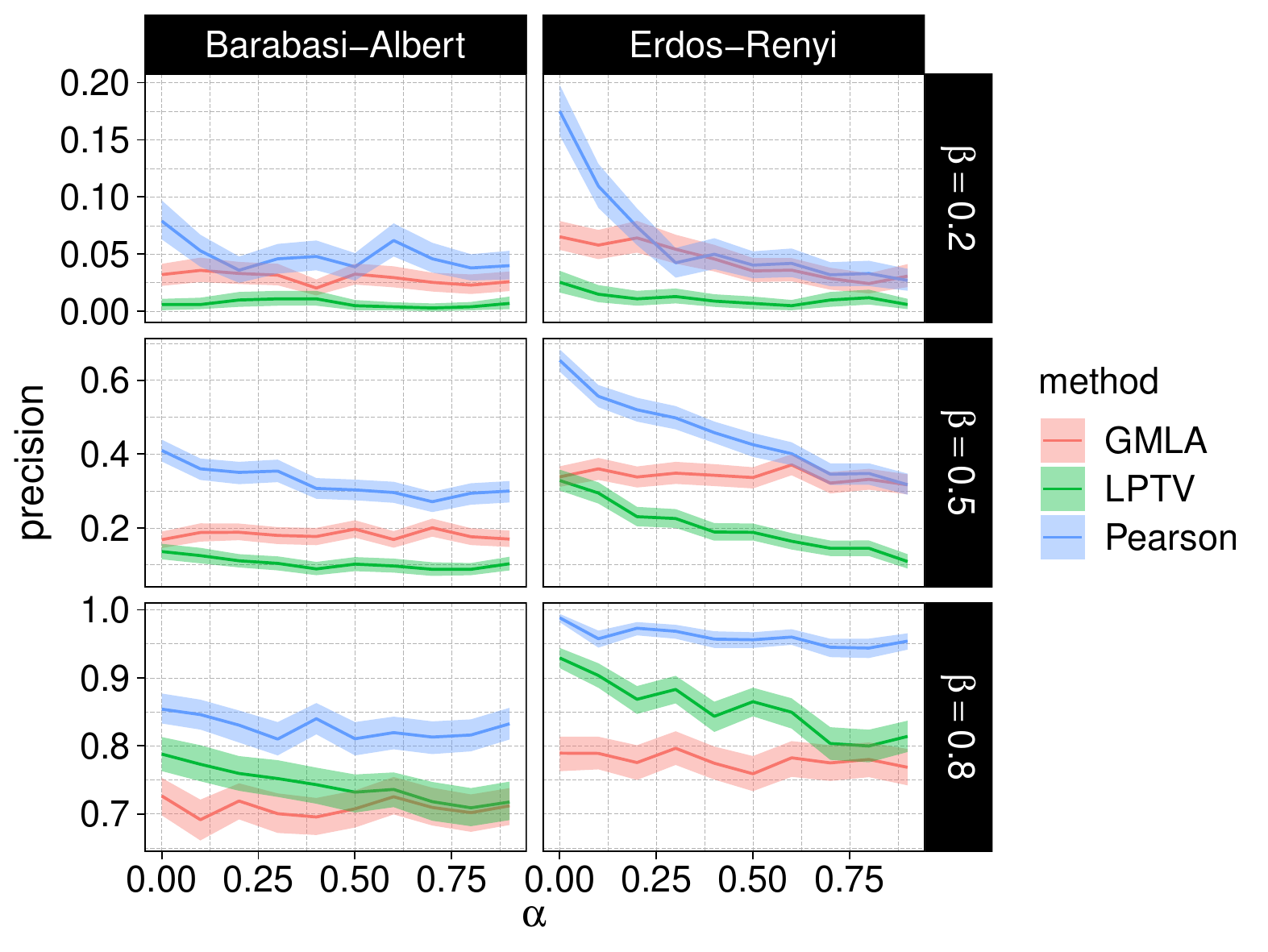}
    \caption{Average precision of rumour source localization in Barab\'{a}si-Albert and Erd\H{o}s–R\'{e}nyi graphs with $n=1000$ and $\langle k \rangle = 8$. 
    Density of observers $\rho = 0.1$.
    Forgetting rate $\gamma = 0$.
    The evaluation metrics are computed from $10^3$ realizations.}
    \label{fig:precision_vs_alpha}
\end{figure}

\begin{figure}[!hbt]
    \centering
    \includegraphics[width=\textwidth]{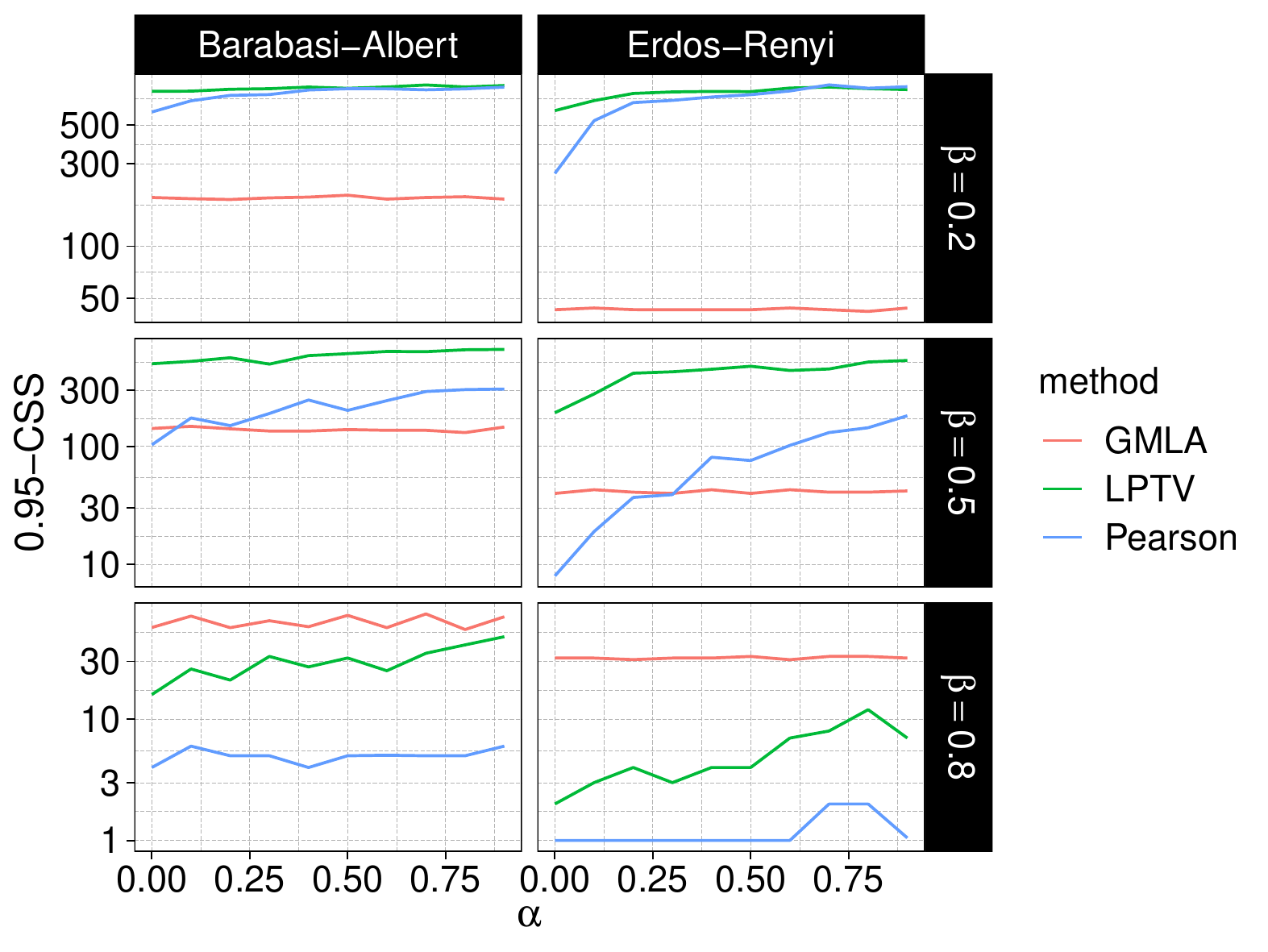}
    \caption{0.95-CSS (lower is better) of rumour source localization in Barab\'{a}si-Albert and Erd\H{o}s–R\'{e}nyi graphs with $n=1000$ and $\langle k \rangle = 8$. 
    Density of observers $\rho = 0.1$.
    Forgetting rate $\gamma = 0$.
    The evaluation metrics are computed from $10^3$ realizations.}
    \label{fig:css_vs_alpha}
\end{figure}

Figures \ref{fig:precision_vs_alpha} and \ref{fig:css_vs_alpha} show the average precision and 0.95-CSS of rumour source localization as a function of parameter $\alpha$, which regulates the level of internal interaction of the spreading process.
A careful analysis of these plots brings up two observations.
Firstly, it can be seen that the self-interaction of signal affects the average precision and 0.95-CSS more in case of Erd\H{o}s–R\'{e}nyi than in Barab\'{a}si-Albert network.
Secondly, each of considered methods for source location reacts slightly differently to additional interactions.
The strongest influence of parameter $\alpha$ on Pearson correlation is seen for low and medium values of spreading rate $\beta$.
On the other hand, the quality of LPTV is affected when spreading rate is medium or high.
At last, GMLA occurs to be the most stable in terms of self-interaction of the signal -- only in case of Erd\H{o}s–R\'{e}nyi graph and $\beta=0.2$ the average precision decreases with the value of parameter $\alpha$.

\section{Discussion}
\label{section-discussion}
In this paper we investigate the problem of the locating the source of self-interacting signal in complex networks.
We use well-known rumour model \cite{Moreno2004, Nekovee2007} as an example of the process with internal interaction.
According to this agent-based model, the infected nodes may interact and discourage each other from gossiping with probability $\alpha$.
We consider the scenario when a single node is the source which starts process of rumour spreading.
The development of an infodemic is observed by the detectors deployed randomly in the nodes of the network. 
Then, we use the information from these observers as an input for three algorithms of source localization: LPTV \cite{Pinto2012}, GMLA \cite{Paluch2018} and Pearson correlation \cite{Xu2019}.
We evaluate the quality of these algorithms in conditions of increasing signal self-coupling.
As expected, the results of numerical simulations show that additional interactions between infected nodes decrease the quality of the source localization.
This effect is stronger in case of Erd\H{o}s–R\'{e}nyi graph
than Barab\'{a}si-Albert model.
This may be due to the fact that epidemic processes are much more explosive in scale-free free networks than in networks with narrow degree distribution. 
This property in general makes the source localization harder, but on the other hand it may diminish the meaning of the internal interactions.
The results also reveal that GMLA is the most resistant to harmful effects of the self-interactions, which is especially visible for medium and high level of stochasticity of the process (when spreading rate $\beta \leq 0.5$).
According to Fig. \ref{fig:css_vs_alpha}, in such cases GMLA provides the highest quality of source localization.
The reason for this may be the fact that GMLA uses only the nearest observers, which are probably much less affected by the interactions between infected nodes, because these contacts become important as the epidemics develops and the number of infected agents increases.

In the present work we consider only one type of self-interactions, which limits the spread of rumours in the system.
It would be worthy to investigate more realistic model, which takes into account positive coupling in signal and the heterogeneity of rumors.
Here we show that the topology of graph also plays a crucial role, therefore the studies with real networks would be another important line of research.

\section*{Acknowledgments}
The work was partially supported by the National Science Centre, Poland Grant No. 2015\-/19\-/B\-/ST6\-/02612.
R.P. was partially supported by the National Science Centre, Poland, agreement No 2019/\-32/\-T/\-ST6/\-00173, and by PLGrid Infrastructure.
J.A.H. was partially supported by the Russian Science Foundation, Agreement No 17-71-30029 with co-financing of Bank Saint Petersburg, Russia.

\typeout{}
\bibliographystyle{ieeetr}
\bibliography{references}

\end{document}